\documentclass[12pt,preprint]{aastex}
\usepackage[english]{babel}
\usepackage{graphicx}
\usepackage{xspace}

\newcommand{\fe}{\snname{2011fe}}

\newcommand{\by}{\snname{2011by}}

\newcommand{\jj}{\snname{2014J}}

\newcommand{\uvwone}{{\it uvw1}\xspace}
\newcommand{\uvwtwo}{{\it uvw2}\xspace}

\newcommand{\wfczero}{F218W\xspace}
\newcommand{\wfcone}{F225W\xspace}
\newcommand{\wfctwo}{F275W\xspace}
\newcommand{\wfcthree}{F336W\xspace}
\newcommand{\wfcBband}{F438W\xspace}
\newcommand{\wfcVband}{F555W\xspace}
\newcommand{\wfcIband}{F814W\xspace}
\newcommand{\wfcsmallB}{F467M\xspace}
\newcommand{\wfcsmallR}{F631N\xspace}
\newcommand{\wfcsmallI}{F845M\xspace}

\newcommand{\Uband}{{\it U}\xspace}
\newcommand{\Bband}{{\it B}\xspace}
\newcommand{\Vband}{{\it V}\xspace}
\newcommand{\Rband}{{\it R}\xspace}
\newcommand{\vband}{{\it v}\xspace}

\newcommand{\iband}{{\it i}\xspace}
\newcommand{\Jband}{{\it J}\xspace}
\newcommand{\Hband}{{\it H}\xspace}

\newcommand{\Ksband}{{\it Ks}\xspace}
\newcommand{\Xband}{{\it X}\xspace}
\newcommand{\EXY}[2]{E(\textrm{#1}-\textrm{#2})}
\newcommand{\EBV}{\EXY{\Bband}{\Vband}}

\newcommand{\snoopy}{SNooPy\xspace}

\newcommand{\eg}{e.g.\xspace}
\newcommand{\sn}{SN\xspace}
\newcommand{\sne}{SNe\xspace}
\newcommand{\snia}{SN~Ia\xspace}
\newcommand{\sneia}{SNe~Ia\xspace}
\newcommand{\snname}[1]{SN\,#1\xspace}

\newcommand{\RV}{R_\mathrm{\Vband}}
\newcommand{\AV}{A_\mathrm{\Vband}}

\newcommand{\swift}{Swift\xspace}
\newcommand{\abu}{Mount Abu\xspace}

\newcommand{\mjdcol}{1}
\newcommand{\phasecol}{2}
\newcommand{\filtcol}{3}
\newcommand{\magone}{4}
\newcommand{\Aone}{5}
\newcommand{\matchcol}{6}
\newcommand{\magtwo}{7}
\newcommand{\Atwo}{8}
\newcommand{\ic}{9}

\newcommand{\goobar}{G14}

\def\lsim{\raise0.3ex\hbox{$<$}\kern-0.75em{\lower0.65ex\hbox{$\sim$}}}
\def\gsim{\raise0.3ex\hbox{$>$}\kern-0.75em{\lower0.65ex\hbox{$\sim$}}}

\shorttitle{The peculiar extinction law of SN~2014J} \shortauthors{Amanullah
  et. al.}  \title{The peculiar extinction law of SN2014J measured
  with {\em The Hubble Space Telescope}}
\author{R.~Amanullah\altaffilmark{1}, A.~Goobar\altaffilmark{1},
  J.~Johansson\altaffilmark{1}, 
  D.~P.~K.~Banerjee\altaffilmark{2}, V.~Venkataraman\altaffilmark{2},
  V.~Joshi\altaffilmark{2}, N.~M.~Ashok\altaffilmark{2},
  Y.~Cao\altaffilmark{3}, M.~M.~Kasliwal\altaffilmark{4},
  S.~R.~Kulkarni\altaffilmark{3} P.~E.~Nugent\altaffilmark{5,6},
  T.~Petrushevska\altaffilmark{1}, V.~Stanishev\altaffilmark{7}  }
\altaffiltext{1}{Oskar Klein Centre, Physics Department, Stockholm
  University, SE 106 91 Stockholm, Sweden} \altaffiltext{2}{Physical
  Research Laboratory, Ahmedabad 380 009, India}
\altaffiltext{3}{Cahill Center for Astrophysics, California Institute
  of Technology, Pasadena, CA 91125, USA}
\altaffiltext{4}{Observatories of the Carnegie Institution for
  Science, 813 Santa Barbara St, Pasadena CA 91101, USA}
\altaffiltext{5}{Department of Astronomy, University of California
  Berkeley, B-20 Hearst Field Annex \# 3411, Berkeley, CA, 94720-3411,
  USA} 
\altaffiltext{6}{Computational Cosmology Center, Computational
  Research Division, Lawrence Berkeley National Laboratory, 1
  Cyclotron\\Road MS 50B-4206, Berkeley, CA, 94720, USA}
\altaffiltext{7}{CENTRA - Centro Multidisciplinar de Astrof\'isica, Instituto 
  Superior T\'ecnico, Av. Rovisco Pais 1, 1049-001 Lisbon, Portugal}
\email{rahman@fysik.su.se}

\begin{document}

\label{firstpage}


\begin{abstract}
  The wavelength-dependence of the extinction of Type Ia \jj in the
  nearby galaxy M82 has been measured using UV to near-IR photometry
  obtained with the Hubble Space Telescope, the Nordic Optical
  Telescope, and the \abu Infrared Telescope. This is the first time
  that the reddening of a \snia is characterized over the full
  wavelength range of 0.2--2$\,\mu$m. A total-to-selective 
  extinction, $R_V\geq3.1$, is ruled out with high
  significance. The best fit at maximum using a Galactic type
  extinction law yields $R_V= 1.4 \pm 0.1$.  The observed reddening of \jj is
  also compatible with 
  a power-law extinction,
  $A_{\lambda}/A_V = \left( {\lambda}/ {\lambda_V} \right)^{p}$ as
  expected from multiple scattering of light, with $p=-2.1 \pm 0.1$.
  After correction for differences in reddening, \jj appears to be
  very similar to \fe over the 14 broad-band filter lightcurves
  used in our study.
 \end{abstract}


\section{Introduction}
The study of the cosmological expansion history using Type~Ia
supernovae (\sneia), of which \jj is the closest in several decades
\citep[][hereafter G14]{2014ApJ...784L..12G} has revolutionized our
picture of the Universe.  The discovery of the accelerating Universe
\citep{1998AJ....116.1009R,1999ApJ...517..565P} has lead to one of the
biggest scientific challenges of our time: probing the nature of {\em
  dark energy} through more accurate measurements of cosmological
distances and the growth of structure in the universe.  \sneia remain
among the best tools to measure distances and as the sample grows both
in numbers and redshift range, special attention is required in addressing
systematic effects.
One important source of uncertainty is the effect of dimming by dust.
In spite of considerable effort, it remains unclear why the color-brightness relation 
for \sne Ia from cosmological fits is significantly different from e.g. dimming by 
interstellar dust with an average $\RV = \AV /\EBV = 3.1$.
In the most recent compilation by
\citet{2014arXiv1401.4064B}, 740 low and high-$z$ \sneia were used to build a Hubble
diagram using the SALT2 lightcurve fitter
\citep{2007AA...466...11G}. Their analysis yields $\beta=3.101 \pm 0.075$, 
which corresponds to $\RV \sim 2$, although the assumed color law in SALT2 differs
from the standard Milky-Way type extinction law \citep{1989ApJ...345..245C}. 

Several cases of $\RV \lsim 2$ has been found in studies of color excesses of local, well-measured, \sneia
\citep[\eg][]{2006AJ....131.1639K,2006MNRAS.369.1880E,2008MNRAS.384..107E,2008A&A...487...19N,2010AJ....139..120F}.
A low value of $\RV$ corresponds to steeper wavelength dependence of the extinction,
especially for shorter wavelengths. In general terms, this reflects the distribution of dust grain 
sizes where a low $\RV$ implies that the light encounters mainly small dust grains.
\citet{2005ApJ...635L..33W} and \citet{2008ApJ...686L.103G} suggest an alternative explanation that
non-standard reddening of \sneia could originate from multiple
scattering of light, e.g., due to a dusty circumstellar medium, 
a scenario that has been inferred for a few \sneia
\citep{2007Sci...317..924P,2009ApJ...693..207B,2012Sci...337..942D}.

A tell-tale signature of reddening through multiple scattering is a power-law dependence for
reddening \citep{2008ApJ...686L.103G}, possibly also accompanied by a
perturbation of the lightcurve shapes \citep{2011ApJ...735...20A} and IR emission from 
heated dust regions \citep{2013MNRAS.431L..43J}.

\jj in the nearby galaxy M\,82 offers a
unique opportunity to study the reddening of a spectroscopically normal 
(\goobar; Marion~et~al. in prep., 2014)
\snia, over an unusually wide wavelength range.   
Hubble Space Telescope (HST)
observations allow us to perform a unique study of color excess in the
optical and near-UV, where the difference between the
extinction models is the largest. Our data-set is complemented by
\Uband\Bband\Vband\Rband\iband observations from the Nordic Optical
Telescope (NOT) and \Jband\Hband\Ksband from the \abu Observatory.

\section{Observations and data}
\subsection{HST/WFC3}
We obtained observations (Program~DD-13621; PI~Goobar) of \jj with HST in the 
four UV broadband filters \wfczero, \wfcone, \wfctwo and \wfcthree for
seven epochs using a total of 7 HST orbits during Cycle~21. In addition to this we also obtained
optical broad, medium and narrow band photometry in filters \wfcsmallB,
\wfcsmallR and \wfcsmallI for visits (1,3) and optical broad-band
photometry using \wfcBband, \wfcVband and \wfcIband for the remaining five
visits.  All observations were obtained with the Wide-Field Camera-3
(WFC3) using the UVIS aperture \texttt{UVIS2-C512C-SUB}.

The data were reduced using the standard reduction pipeline and
calibrated through CALWF3 as integrated into the HST archive.  The flat-fielded
images were corrected for charge transfer inefficiencies at the pixel
level\footnote{J.~Anderson, private communication} and photometry was
carried out on the individual images following the guideline
from the WFC3 Data Handbook.  The individual flat-fielded images were 
multiplied with the correcting pixel area
map\footnote{\texttt{http://www.stsci.edu/hst/wfc3/pam/pixel\_area\_maps}}
following Sec.~7 of the WFC3 Data Handbook.
The \sn flux could be measured 
on all images using an aperture with radius 0.2$\,''$. Host contamination is negligible at the SN position and
the statistical uncertainties were estimated assuming Poisson noise of the signal together with the readout noise.
The resulting photometry is presented in Tab.~\ref{tb:phot}.


\begin{table*}
  \begin{center}
    \begin{tabular}{r r c c c c c c c c c}
      \hline\hline
      {\small MJD} & 
      {\small Phase} &
      {\small Filter} & 
      {\small Mag} & 
      {\small $A_X$} &
      {\small Match} &
      {\small \Vband} &
      {\small $\AV$} &
      {\small 2011fe}\\
      \multicolumn{1}{c}{(\small\mjdcol)} & 
      \multicolumn{1}{c}{(\small\phasecol)} & 
      (\small\filtcol) & 
      (\small\magone) & 
      (\small\Aone) & 
      ({\small\matchcol}) &  
      (\small\magtwo) & 
      (\small\Atwo) & 
      (\small\ic)\\
      \hline
{\small 56685.0} & {\small -3.6} & F218W & {\small 18.18(0.01)} & {\small 0.20} & D & {\small 10.97(0.02)} & {\small 0.15} & {\small $-3.22$}\\
{\small 56688.8} & {\small -0.2} & F218W & {\small 18.03(0.01)} & {\small 0.20} & M & {\small 10.68(0.02)} & {\small 0.15} & {\small $-3.13$}\\
{\small 56692.1} & {\small 2.9} & F218W & {\small 18.03(0.01)} & {\small 0.20} & M & {\small 10.67(0.02)} & {\small 0.15} & {\small $-3.14$}\\
{\small 56697.0} & {\small 7.3} & F218W & {\small 18.35(0.02)} & {\small 0.20} & D & {\small 10.81(0.02)} & {\small 0.15} & {\small $-3.53$}\\
{\small 56702.9} & {\small 12.7} & F218W & {\small 18.88(0.01)} & {\small 0.19} & M & {\small 11.02(0.02)} & {\small 0.15} & {\small $-4.00$}\\
{\small 56713.7} & {\small 22.6} & F218W & {\small 19.85(0.03)} & {\small 0.17} & M & {\small 11.55(0.02)} & {\small 0.15} & {\small $-4.43$}\\
	\multicolumn{9}{c}{$\cdots$}\\
      \hline
    \end{tabular}
    \caption{The measured photometry of \jj from HST/WFC3, NOT/ALFOSC and the \abu infrared telescope.  
       All magnitudes are in the natural Vega system.  The rest-frame magnitude can be obtained as 
       $(\magone) - (\Aone)$ where
       column $(\Aone)$ and $(\Atwo)$ are the Galactic extinctions for the two bands respectively.
       Column $(\phasecol)$  shows the effective, lightcurve width corrected, phase, while column $(\matchcol)$ 
       specifies if the \Vband magnitude was measured for the same epoch (D) or if it was calculated
       using the \snoopy model (M). In the latter case the mean error of the data used for the fit was adopted as the
       uncertainty of the magnitude.
       The corresponding synthesized color of \fe is shown in column (\ic).  {\em The full table is available in electronic 
       format online}.
      \label{tb:phot}}
  \end{center}
\end{table*}

\begin{figure}
  \begin{center}
    \includegraphics[width=\columnwidth]{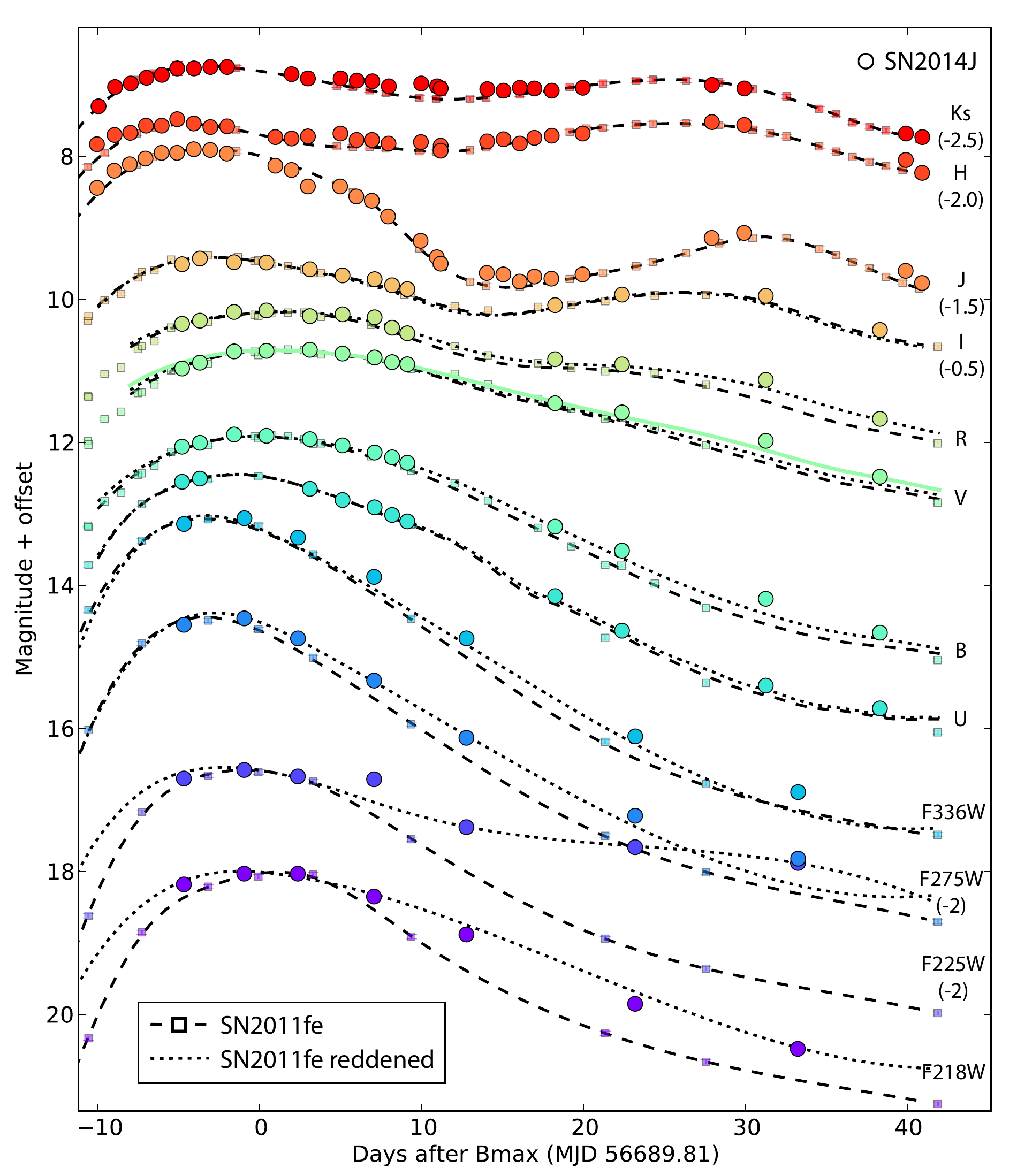}
    \caption{Lightcurves for all passbands used in this analysis.  For \Vband-band we also over plot (solid, green line) 
       the fitted model from \snoopy. The black lines are fits to synthetic photometry of \fe spectra (dashed) and 
       spectra reddened with the best fit FTZ model to \jj.
       \label{fig:lc}}
  \end{center}
\end{figure}

\subsection{Optical and near-IR data}
The \Uband\Bband\Vband\Rband\iband data were obtained with the NOT
(Program 48-004; PI~Amanullah). The data were reduced with standard
IRAF routines, using the QUBA pipeline \citep[see][for
details]{2011MNRAS.416.3138V}. The magnitudes are measured with a
PSF-fitting technique (using daophot) and calibrated to the Landolt
system. 


NIR observations in the Mauna Kea Observatory \Jband\Hband\Ksband filters were 
carried out with the Mount Abu 1.2\,m  Infrared telescope. Aperture photometry of
the sky-subtracted frames was done using IRAF.  The nearby star 
2MASS J09553494+6938552, which registers simultaneously with \jj in
the same field, was used for calibration. Results were cross-checked with other 2MASS stars in
the field and found to agree within 5\,\%.  We adopt this 
as a systematic uncertainty on the NIR photometry.


All lightcurves\footnote{All tables and figures are available at {\tt http://www.fysik.su.se/{\textasciitilde}rahman/SN2014J/}} 
are presented in Table 1 and Fig.~\ref{fig:lc} where we
also show a fitted model to the \Vband-band using \snoopy
\citep{2011AJ....141...19B}.

\section{Color excess}
\subsection{Intrinsic SN Ia colors}\label{sec:intrinsic}
In order to study the reddening of \jj, the colors for a pristine, unreddened, \snia must be known.  Further, we need
a color template that includes both the wavelengths and epochs covered by the observations presented in this work.  

As described in \goobar, the early spectral evolution of \jj and \fe
in the nearby spiral galaxy M\,101 is remarkably similar.  The only
difference being that \jj shows overall higher photospheric velocities. 
\fe has been observed over a broad wavelength
range from the UV \citep{2012ApJ...753...22B,2014MNRAS.439.1959M},
through the optical \citep[\eg][]{2013A&A...554A..27P}, to the near-IR
\citep{2012ApJ...754...19M,2013ApJ...766...72H}.
The similarity to \jj together with the low Galactic and host galaxy reddening, 
$\EBV_{\rm MW} = 0.011 \pm 0.002$ and $\EBV_{\rm host} =
0.014 \pm 0.002$~mag \citep{2013A&A...549A..62P}, makes \fe the best
template we can derive of the unreddened spectral energy distribution (SED) of \jj. 

We use the spectral series from \citet{2014MNRAS.439.1959M} and
\citet{2013A&A...554A..27P}, corrected for Galactic extinction using  
\citet[][FTZ from hereon]{1999PASP..111...63F} with $\RV=3.1$, to compute
synthetic colors between in the WFC3 and NOT bands in which \jj was observed.  
The effective HST filters were obtained from \textsc{synphot/stsdas}, while we used 
modified versions of the public effective NOT filters.  We further use the \fe lightcurves from 
\citet{2012ApJ...754...19M} as an unreddened NIR template of \jj. All lightcurves are shown
in Fig.~\ref{fig:lc} (dashed curves) where they have been shifted to overlap with the corresponding \jj 
photometry at maximum. Smoothed splines are fitted using \snoopy to each individual band
to create a pristine lightcurve template.  

As seen in
Fig.~\ref{fig:lc}, the NIR lightcurves of \fe provide an excellent
description of the corresponding bands of \jj, while this is not
the case for the bluer bands. At these wavelengths \fe
appears to both rise and fall faster than \jj, and the difference
between the two objects increases with shorter wavelengths.  As will be
argued in Sec.~\ref{sec:law} this is partially an effect that
stems from the fact that broadband observations of \jj are effectively
probing longer wavelengths than the corresponding data of \fe due to
the significant extinction. Taking this effect into account leads to the dotted lines 
for the reddened SED of \fe in Fig.~\ref{fig:lc}.

The spectral series will also be used in the analysis to calculate the expected 
extinction in each passband for a given extinction law.  
The \citet{2014MNRAS.439.1959M} dataset extend out to $\sim2\,\mu$m until 
phase $+9$.  Since this does not cover the entire phase-range of our study 
we extended this spectral series using the template from 
\citet{2007ApJ...663.1187H,2013ApJ...766...72H} for phases past $+9$. 

In order to compare \jj with \fe, we
also need an estimate of the uncertainty within which we would expect
the broadband colors of two \sneia to agree in the absence of
extinction.
\citet{2010AJ....139..120F} studied the intrinsic optical and near-IR colors of
\sneia close to lightcurve maximum, and found dispersions in the range $0.06$--$0.14$ mag, after correcting
for lightcurve shape.  We conservatively
adopt a dispersion of 0.15~mag (the worst case above) for all colors that only include
the optical and NIR bands.

Further, \citet{2010ApJ...721.1627M} presented an extensive study of
the UV$-$\Vband dispersion based on observations of 12~\sneia with the
\swift satellite. For their low-extinction ($\EBV<0.2$) sample they
derive dispersions of $0.1$ and $0.25$~mag between $-12$ and
$+12$~days relative \Bband-band maximum for the \uvwone$ - $\vband and
\uvwtwo$ - $\vband colors respectively.
We adopt a dispersion of $0.35$~mag for the colors that
involve \wfczero and \wfcone and $0.25$~mag  for \wfctwo. For the $\Vband-\wfcthree$
dispersion we adopt the same value as the optical range, i.e.,~$0.15$~mag. 
Since UV observations of \sneia are scarce it is difficult to fully assess
the differences among supernovae at the shortest wavelengths considered here. 
\citet{2013ApJ...769L...1F} argued that although \by was a
spectral ``twin'' to \fe in the optical, it exhibited a different
behavior in the near-UV. We have therefore checked how our estimate
of the color excess of \jj would differ under the assumption that it
is a better match to \by instead of \fe. The offsets at lightcurve
maximum are $\Delta E(V-\wfcone)=0.35$ mag and $\Delta E(V-\wfctwo) = 0.03$ mag, 
i.e., compatible with our estimate of the intrinsic color scatter.


\begin{figure*}
  \begin{center}
    \includegraphics[width=\textwidth]{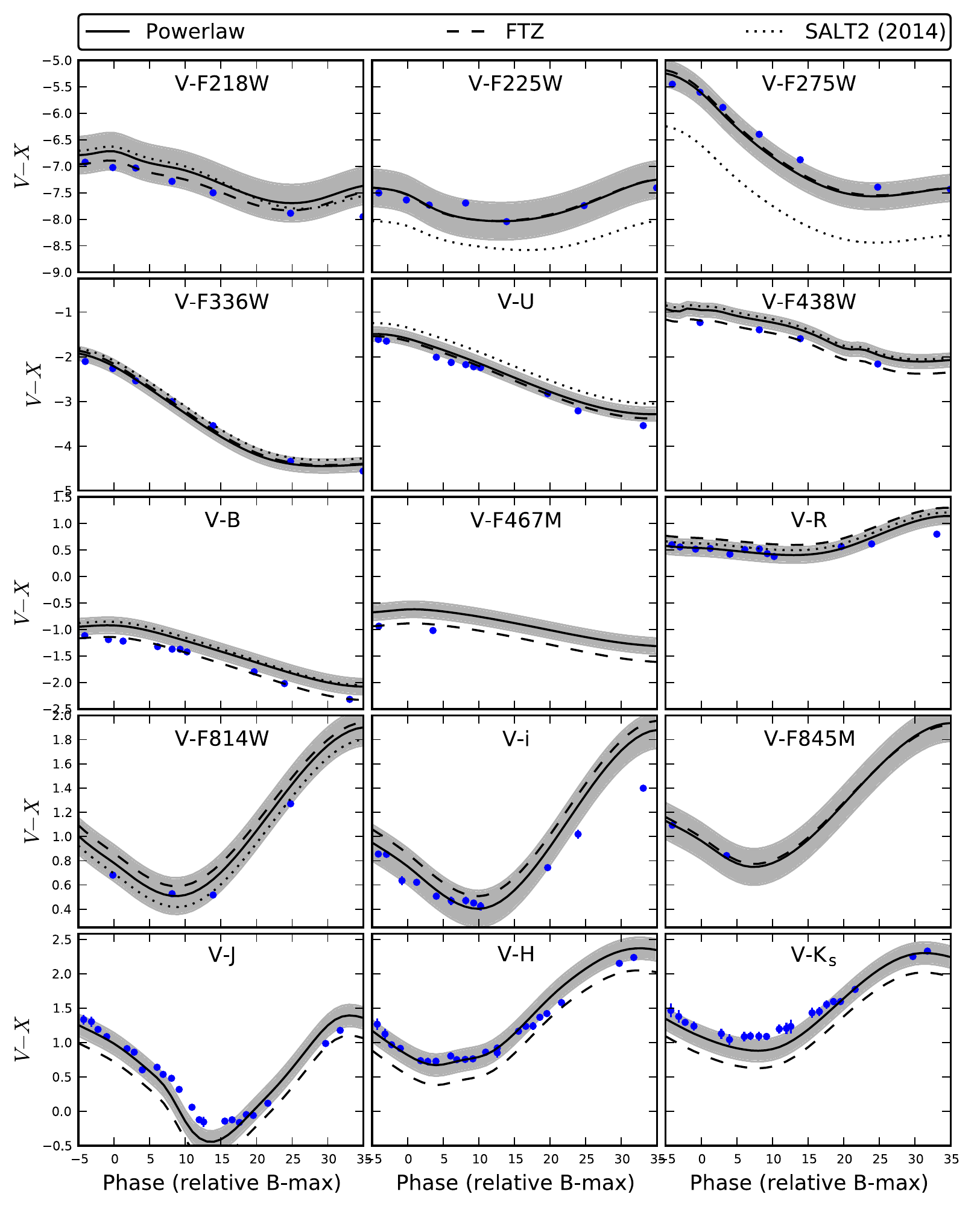}
    \caption{The measured colors (blue points) for the UV, optical and NIR bands.  Also shown are the best
    	fitted extinction laws in the $[-5,+35]$ range together with the corresponding predicted colors of \fe.  
	The grey band shows the adopted intrinsic dispersion of each color plotted with respect to the 
	power-law fit.  \label{fig:color}}
  \end{center}
\end{figure*}

\subsection{Color excesses of \jj}
In this work we study the color excesses, $E(V-X)$, of all photometric bands with
respect to the \Vband-band.  For each photometric
observation in Table~\ref{tb:phot} we also list the corresponding
\Vband magnitude.  If the \sn was observed in both bands within
12\,hours we use the observed \Vband for the corresponding epoch, but
when this was not the case we use the fitted \Vband-band \snoopy model
shown in Fig.~\ref{fig:lc} to calculate the color.

For each epoch we also present the calculated Galactic reddening
correction.  Unlike G14, we use the Galactic extinction towards M\,82 
from \citet{2009ApJS..183...67D}. They argue that the
estimates from the dust maps of \citet{1998ApJ...500..525S} are
contaminated by M\,82 itself, and derived $E(B-V)_{MW}=0.06$ from the
study of neighboring patches.

We also calculate the corresponding color of \fe, shown in the last
column of Table~\ref{tb:phot}, from the lightcurves described
above. The color excess, $E(\Vband_n - \Xband_n)$, between 
the \Vband-band and some other band \Xband, at an epoch $n$, can then
be obtained under the assumption that the two \sne had nearly identical color evolution.
Since the differences in {\it K}-corrections are negligible for the two very nearby \sne,
the color excess is calculated as the difference between the $\Vband_n-\Xband_n$ color, corrected 
for Galactic extinction, and the corresponding color of \fe, i.e.
\begin{equation}
E(\Vband_n - \Xband_n) =  \left[(V_n -A_{V_n}) - (X_n - A_{X_n})\right] - (V_{n}^{\rm 11fe} - X_{n}^{\rm 11fe})\,,
\end{equation}
The first term on the right hand side, the extinction corrected color of \jj, is plotted in Fig.~\ref{fig:color}
together with the best fit colors derived from the reddened SED, as described in the next section.
In Fig.~\ref{fig:extinctionlaw}, $E(\Vband_n - \Xband_n)$ is plotted using data around maximum light.

\section{Fitting extinction laws}\label{sec:law}
An extinction law, $\xi(\lambda;\bar{p})$, where $\bar{p}$ are free parameters, can be fitted by minimizing
\newcommand{\fesed}{S_{\mathrm{11fe}}}
\begin{equation}
	\chi^2 = \sum_{X}\sum_{n} \frac{\left[ (E(\Vband_n-\Xband_n) - (A_{V_n} - A_{X_n})\right]^2}{\sigma_{V_nX_n}^2}\, .
\label{eq:chi2}
\end{equation}
Here $E(\Vband_n-\Xband_n)$ is the measured color excess as described above and $A_{V_n} - A_{X_n}$, the corresponding model color excess,
\begin{equation}
	A_{X_n} = -2.5\log_{10}\left(\frac{\int \xi(\lambda;\bar{p})T_X(\lambda)\fesed(\lambda;n)\lambda\,d\lambda}{%
		\int T_X(\lambda)\fesed(\lambda;n)\lambda\,d\lambda}\right)
\label{eq:ax}
\end{equation}
can be calculated from the filter transmission, $T_X(\lambda)$, and the SED of \fe, 
$\fesed(\lambda;n)$, of the given epoch $n$.

The uncertainties, $\sigma_{V_nX_n}^2$, include the measurement errors
shown in Table~\ref{tb:phot} but is dominated by the
adopted intrinsic \snia color uncertainties.  If a \Vband-band measurement is 
included in constructing two different colors for a given epoch, then the 
contribution from the \Vband uncertainty is treated as fully correlated.

In this work we ignore both calibration uncertainties (except for NIR) and 
systematic errors due to e.g. Galactic extinction correction.  The reason being
that these will be correlated between epochs and are negligible
in comparison to the intrinsic color uncertainty, that we also assume
to be fully correlated. This will put equal weights to all colors, independently 
of the number of data points obtained in each band.

We further assume that the different colors are uncorrelated in our final
analysis.  We have tried inducing different correlations between the
colors, but this had a minor impact on the conclusions from the fits.

We have fitted three different extinction laws: a MW-like extinction
law as parametrized by \citet[][FTZ]{1999PASP..111...63F}, the
SALT2 law in the version used in
\citet{2014arXiv1401.4064B}, and a power-law parametrization, 
$A_\lambda/A_v = (\lambda/\lambda_V)^p$, 
shown to be a good approximation for multiple scattering scenarios
 \citep{2008ApJ...686L.103G}.  Each law was fitted
using three different epoch ranges: $[-5, +5]$ days, i.e., around peak 
brightness, the tail: $[+25,+35]$, and the full epoch range of the HST 
observations,  $[-5, +35]$ days since \Bband-band maximum. The 
FTZ and power-law were fitted using all filters, while the SALT2 law 
is only defined in the wavelength range $2000$--$7990\,$\AA.

The \sn colors are compared at the effective phases, computed for 
both \fe and \jj by subtracting the date of \Bband-max, $55814.5$ and 
$56689.2$ respectively, and dividing by the \iband lightcurve stretches (from \snoopy)
$0.96$ and $1.14$ respectively. For each law we carry out 
the fit iteratively to all bands except F555W and F631N, The former completely
includes our reference band and the latter is too narrow for our assumption
of \jj and \fe being comparable to hold.

\begin{figure}
  \begin{center}
    \includegraphics[width=\columnwidth]{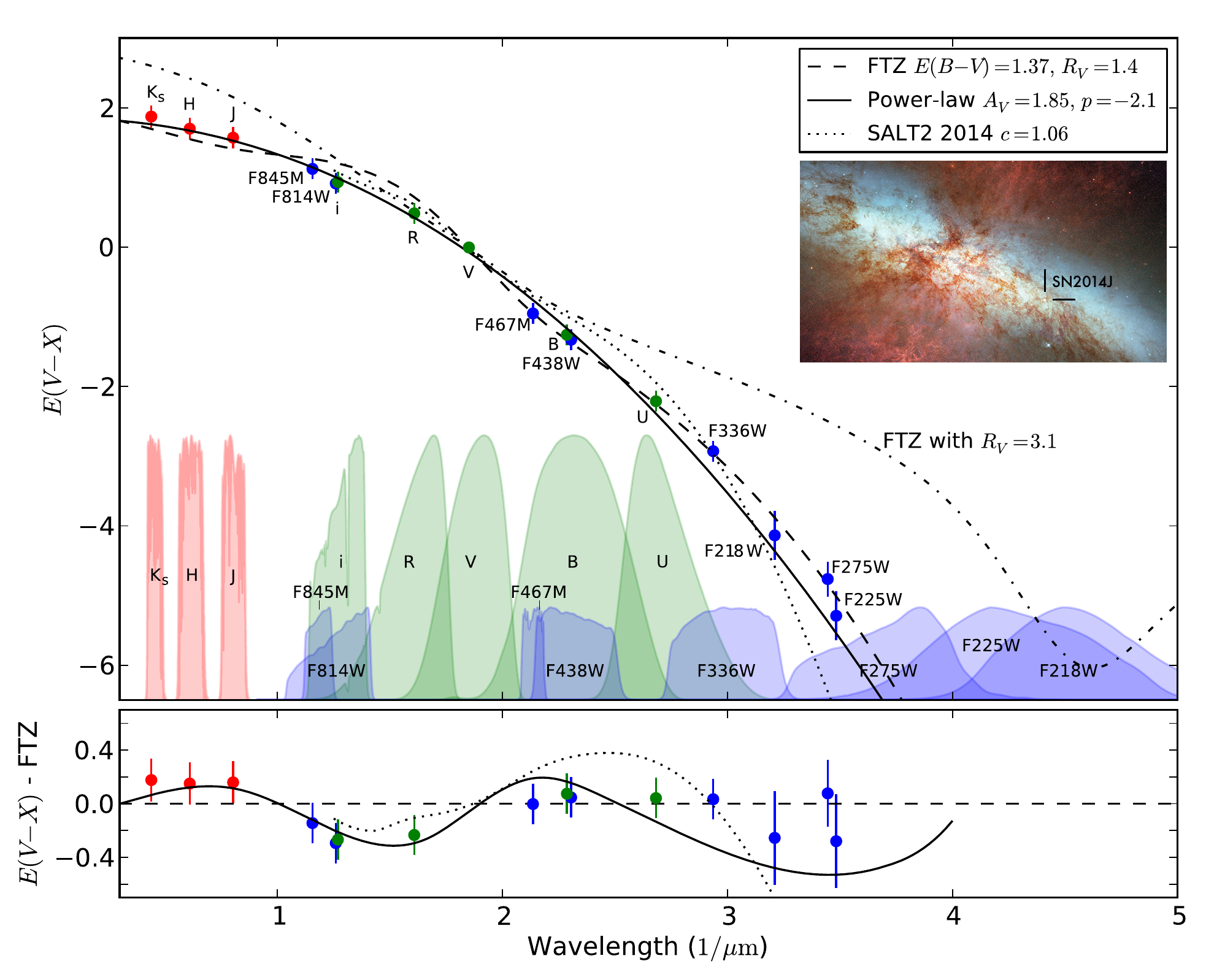}
    \caption{The {\em upper} panel shows the average color excesses between $-5$ and $+5$ days 
      from \Bband-maximum.  
      Blue, green and red points are measured with HST, NOT and the
      Mount Abu Infrared Telescope, respectively, while the corresponding effective filter transmissions are plotted,
      in linear scale, at the bottom of the panel.  The same data points have been plotted as residuals with respect to the
      best fitted FTZ law in the {\em lower} panel. For the UV-filters the effective wavelengths are significantly redder
      than the central wavelengths due to the steepness of the reddened \sn spectrum.\label{fig:extinctionlaw}}
  \end{center}
\end{figure}

We first calculate the Galactic extinction in each band assuming the pristine 
SED of \fe. Then, we carry out the fit by minimizing Eq.~(\ref{eq:chi2}). 
The fitted extinction law is then applied to the SED of \fe and the 
Galactic extinction is recalculated before refitting
the law. The procedure is repeated until the value of the 
fitted parameters change by less than 1\,\% between iterations.  
The results of the fits are shown in Table~\ref{tb:results}, and the best 
fits to the full range are also shown in Fig.~\ref{fig:color} while the best
fit around maximum is shown in Fig.~\ref{fig:extinctionlaw}. In this figure we
also present the best fitted FTZ model with $\RV$ fixed to $\RV=3.1$,
which is clearly excluded by the data. Our best fit FTZ values are 
$E(B-V) = 1.37 \pm 0.03$, $\RV= 1.4 \pm 0.1$. We find that our data are also
compatible with the power-law model with $A_V = 1.85\pm0.11$ and $p= -2.1\pm0.1$. 
Finally, we conclude that the SALT2 model provides a somewhat poorer fit description with 
$c = 1.06\pm0.04$. These findings can be compared with the measured global 
extinction in M\,82 by \citet{2014MNRAS.440..150H}. Unlike our result for the line of 
sight of \jj, they conclude that FTZ with $\RV=3.1$ provides a good description of 
the colors of the galaxy based on stellar modelling. However, for the dust in the 
superwind, they too conclude that a power-law relation provides the best fit, albeit
 with $p=-1.53 \pm 0.17$. 

As a consistency check, we recalculate the synthetic lightcurves of \fe
using the best fitted FTZ law to redden the spectra.
The result is plotted in Fig.~\ref{fig:lc} as dotted lines for each band.  
For the redder bands, these reddened lightcurves are similar to the 
original, but for the blue bands, and in particular in the UV, the  
lightcurves are significantly broader, and show a similar decline 
as the observed data of \jj.  We take this as yet another confirmation 
that \fe indeed is very similar to \jj and therefore suitable to use as 
reference for the extinction study presented here.  We further conclude 
\citep[and references therein]{2002PASP..114..803N} that the difference in lightcurve width between 
the original lightcurves in the bluer bands mainly stems from the fact that we 
are probing redder effective wavelengths for \jj than for \fe when comparing the
same passbands.

This also has implications for Fig.~\ref{fig:extinctionlaw}.  Here we
have calculated the weighted average for each color taking the full
covariance into account.  However, the color excess is calculated by
comparing magnitudes of a reddened and an unreddened source.  
The two magnitudes will correspond to different effective wavelengths, 
and the broader the filter, and the steeper the spectrum, the larger the difference
of the two effective wavelengths will become.  For Fig.~\ref{fig:extinctionlaw} we 
allowed the wavelength to shift, with respect to the average effective wavelength
using the FTZ law, until the residuals of the color excess match the weighted 
average residual from the fit.  For the bluest  \wfczero and \wfcone the shift 
becomes $330\,$\AA\ and $240\,$\AA\ respectively. Both of these filters suffer 
from minor red leaks, e.g. for \wfczero 0.3\,\% of the light comes from wavelengths 
redder than $4000\,$\AA.  On the other hand, a \snia at maximum with the 
reddening of \jj will typically be $\sim6$~orders of magnitude brighter at $4000\,$\AA\  
compared to the central wavelength of the \wfczero filter.  The significant
shift towards redder wavelengths for these filters is in other words
not surprising.

\begin{table*}
	\begin{center}
	\begin{tabular}{l c c c c c c c c}
	\hline\hline\\[-1.5ex]
\multicolumn{1}{c}{} & \multicolumn{3}{c}{FTZ} & \multicolumn{3}{c}{Power-law: $A_\lambda = A_V \left(\lambda/\lambda_V\right)^p$ } & \multicolumn{2}{c}{SALT2 (2014)}\\[1.5ex]
 \multicolumn{1}{c}{Phase} & $\EBV$ & $\RV$ & $\chi^2/\mathrm{dof}$ & $\AV$ & $p$ & $\chi^2/\mathrm{dof}$ & $c$ & $\chi^2/\mathrm{dof}$\\
\hline
$[-5,+5]$ & {\small $1.37(0.03)$} & {\small $1.4(0.1)$} & {\small 1.1} & {\small $1.85(0.11)$} & {\small $-2.1(0.1)$} & {\small 1.1} & {\small $1.06(0.04)$} & {\small 5.3}\\
$[+25,+35]$ & {\small $1.33(0.04)$} & {\small $1.3(0.1)$} & {\small 1.9} & {\small $1.52(0.11)$} & {\small $-2.4(0.1)$} & {\small 2.2} & {\small $1.10(0.05)$} & {\small 6.9}\\
$[-5,+35]$ & {\small $1.29(0.02)$} & {\small $1.3(0.1)$} & {\small 3.3} & {\small $1.77(0.10)$} & {\small $-2.1(0.1)$} & {\small 2.3} & {\small $1.00(0.01)$} & {\small 5.2}\\
	\hline\hline
	\end{tabular}
	\end{center}
	\caption{The best fitted parameters for each reddening law to the broad band filters.  Quoted errors
		are the uncertainties from the $\chi^2$ fit.\label{tb:results}}
\end{table*}

\section{Conclusions}
We present results from fitting three extinction laws to observations
of \jj in 16~photometric bands spanning the wavelength range
$0.2$--$2\,\mu$m between phases $-5$ and $+35$ days with respect to
\Bband-maximum. We find a remarkably consistent picture with reddening
law fits only involving two free parameters. Once reddening is
accounted for, the similarity between the multi-color lightcurves of
\jj and \fe is striking.

We measure an overall steep extinction law with a total-to-selective
extinction value $\RV$ at maximum of $\RV=1.4\pm0.1$ for a MW-like
extinction law.  We also note that fitted extinction laws are
consistent when fitted separately around maximum and using the full
phase-range.

Although the fits slightly disfavor the empirically derived SALT2
color law for \snia, in comparison to a MW-like extinction law as
parametrized by FTZ with a low $\RV$, conclusions should be drawn
cautiously.  SALT2 has not been specifically trained for the near-UV
region considered here. Also, there is no prediction for the
near-IR. Intriguingly, power-law extinction proposed by
\citet{2008ApJ...686L.103G} as a model for multiple scattering of
light provides a good description of the reddening of \jj.


Increasing this sample is crucial to understand the possible diversity in 
reddening of \sneia used to measure the expansion history of the Universe.

\acknowledgments 
We would like to thank Denise Taylor, Claus Leitherer and John Mackenty
at Space Telescope Science Institute for advising and assisting us in
carrying out this program.
RA and AG acknowledge support from the Swedish Research
Council and the Swedish Space Board.  MMK acknowledges
support from the Hubble Fellowship and Carnegie-Princeton Fellowship.
Observations made with the Hubble Space Telescope, the Nordic 
Optical Telescope, operated by the Nordic Optical Telescope Scientific Association at the
Observatorio del Roque de los Muchachos, La Palma, Spain
and the Mount Abu 1.2m Infrared telescope, India. STSDAS is a product of 
the Space Telescope Science Institute, which is operated by AURA for NASA.
V.S. acknowledges support from Funda\c{c}\~{a}o para a Ci\^{e}ncia 
e a Tecnologia (Ci\^{e}ncia 2008) and grant PTDC/CTE-AST/112582/2009.

\appendix

\bibliographystyle{apj}

\label{lastpage}
\end{document}